\documentclass[]{aa}
\usepackage{epsfig} 
\usepackage{graphicx}
\usepackage{amssymb}
\usepackage{natbib}

\begin{document}

  \title{On the maximum efficiency of the propeller  
mass-ejection mechanism}


\author{M. Falanga\inst{1}\fnmsep\thanks{\email{mfalanga@cea.fr}} 
\and E. Bozzo\inst{2,3}
\and L. Stella\inst{2}
\and L. Burderi\inst{4}
\and T. Di Salvo\inst{5}
\and R. Perna \inst{6}
}

\offprints{M. Falanga}
\titlerunning{Propeller Efficiency}  
\authorrunning{M. Falanga, E. Bozzo, L. Stella  et al.}

\institute{CEA Saclay, DSM/DAPNIA/Service d'Astrophysique (CNRS FRE
2591), F-91191, Gif sur Yvette, France 
\and
INAF - Osservatorio Astronomico di Roma, Via Frascati 33,
00044 Rome, Italy
\and
Dipartimento di Fisica - Universit\`a di Roma "Tor Vergata", via
della Ricerca Scientifica 1, 00133 Rome, Italy
\and
Universit\`a degli studi di Cagliari, Dipartimento di Fisica, SP
Monserrato-Sestu, KM 0.7, 09042 Monserrato, Italy
\and
Department of Astrophysical and Planetary Sciences, University of
Colorado, Boulder, CO, 80309, USA
             }
   \date{}

   \abstract
{}
{We derive simple estimates 
of the maximum efficiency with which matter can be ejected 
by the propeller mechanism in disk-fed, rotating magnetic neutron 
stars. Some binary evolution scenarios envisage that this mechanism 
is responsible for expelling to infinity the mass 
inflowing at a low rate from the companion star,  
therefore limiting the total amount of  
mass that can be accreted by the neutron star.} 
{We demonstrate that, for typical neutron star parameters, 
a maximum of $\eta_{\rm pro} < 5.7
 (P_{-3})^{1/3}$ times more matter 
than accreted can be expelled through the 
propeller mechanism at the expenses of the neutron star rotational energy 
($P_{-3}$ is the NS spin period in unit of $10^{-3}$~s). Approaching
this value, however, would require a great deal of fine tuning in the
system parameters and the properties of the interaction of matter and
magnetic field at the magnetospheric boundary.} 
{We conclude that 
some other mechanism must be invoked in order to prevent that too 
much mass accretes onto the neutron stars of some 
low mass X-ray binaries.}
   {} 

\keywords{accretion: disk  - stars: close binaries -stars: neutron
   -pulsars: general }

   \maketitle

\section{Introduction}
\label{sec:intro} 

One of the possible endpoints of the evolution of the compact object 
in a low-mass X-ray binary (LMXB) is a millisecond pulsar (MSP), 
a rapidly spinning neutron star (NS) with weak surface magnetic field
($\sim 10^{8}-10^{9}$ G).  
Such NSs are thought to be spun-up to millisecond periods
by the accretion of mass and angular momentum from matter in a disk formed 
through Roche-lobe overflow of a low-mass companion star
\citep[for a review see e.g.,][]{BvdH}. After the accretion phase has ended,
the NS may turn on as a radio MSP. 
In recent years, the prediction that LMXB host NSs with spin 
periods in the ms range has been  
confirmed through the discovery of 300 -- 600 Hz nearly-coherent
oscillations during type I X-ray bursts from some 10 
LMXBs of the Atoll class \citep[e.g.,][]{SZS}.
This was followed by the detection of coherent X-ray pulsations 
in the persistent emission of seven low luminosity transients of the same
class. The spin frequencies lie between 180 and
600 Hz and the orbital periods between 40 min and 5 hr
\citep[see review by][]{Wijn05}. For the first time, the predicted
decrease of the NS spin period during accretion was measured in the
accreting MSP IGR J00291+5934 \citep{fal05,burd06}. This provided a strong
confirmation of the theory of 'recycled' pulsars in which the old
NS in LMXBs become millisecond radio pulsars through
spin-up by transfer of angular momentum by the accreting material.

All accreting MSPs are X-ray transient systems.
They spend most of the time in a quiescent phase, with X-ray luminosities
of the order of $10^{31}-10^{32}$ erg s$^{-1}$, and occasionally show 
weeks to months long outbursts, 
reaching X-ray luminosities of $10^{36}-10^{37}$ erg s$^{-1}$,
during which coherent X-ray pulsations are observed. The magnetic fields
inferred for these NS are of order $10^8-10^9$ G
\citep[e.g.,][]{Camp,PC99,DB03}. These findings added a great deal of
confidence in the likely connection between accreting NS and
millisecond radio pulsars. 

Models for the evolution of LMXBs as progenitors of radio MSPs still 
involve significant assumptions about the transfer of mass from one 
star to the other and the role of dynamically influent NS magnetic
fields \citep{PRP}. The MSP formation depends on the amount of matter
accreted from the companion which, in turn, is likely limited by
mass ejection from the binary. 
\citet{CSTa}, following a full GR treatment, show that a NS of 1.4
M$_{\odot}$ has to accrete $\sim 0.2-0.5$ M$_{\odot}$ to reach periods
in the range 0.6--1.5 ms \citep[see also e.g.,][]{b99}. Most donor
stars in systems hosting recycled radio pulsars must have lost 
most of their mass during their evolution. They now appear as
white or brown dwarfs of mass $\sim 0.15-0.30$ M$_{\odot}$ \citep{TKR},
the progenitors of which were likely stars of $\sim
1.0-2.0$ M$_{\odot}$ \citep[e.g.,][]{BKW,TS}. 
Hence, a crucial issue for evolutionary models for  
MSPs is which fraction of mass lost by the
companion (M$_{\rm lost} \sim 0.7-1.8$ M$_{\odot}$)
effectively accretes onto the NS. 
Mass ejection must be efficient, at least in some cases, to avoid that accretion of
a large amount of mass onto the NS could induce collapse to a black hole. 
For most equations of state of ultra-dense matter, this is expected 
to occur for NS masses above $\sim2$ M$_\odot$ \citep{CSTb}.
Different models have been proposed to explain how mass can be 
ejected from binary system, 
thus limiting the amount of matter that is accreted by the NS 
to about 0.2--0.4 M$_{\odot}$ \citep{IS,wr85,romanova05,ug}.

In this work we use simple arguments to derive an upper limit 
on the efficiency of the propeller mass-ejection
mechanism. In section \ref{propeller} we consider
accretion and ejection of infinitesimally small amounts of 
matter, the case relevant to NS X-ray transient.
Since in this case the NS parameters do not change significantly
between each accretion/ejection cycle an analitic expression for the 
propeller efficiency can be derived. 
In section \ref{discussion} we extend this limit to the 
case of accretion and ejection of finite amounts of matter.

\section{Ejector-Propeller efficiency}
\label{propeller}
Matter transferred from the companion star via Roche-Lobe overflow 
possesses enough specific angular momentum that an accretion disk forms 
around the accreting NS. The disk is truncated
at the magnetospheric radius, $R_{\rm m}$. Different calculations,
carried out by adopting different models and/or assumptions,  
produce only slightly different estimates of the value of $R_{\rm m}$ 
\citep[a factor of 2--3 at the most, e.g.,][]{RPS,GL79,W87}.

The value of $R_{\rm m}$ as a function of the system parameters 
is only of marginal relevance to the present paper. Therefore 
we consider here the value inferred from simple theory 
\citep[see e.g., Eq. (4) of][]{LP}; that is
\begin{equation}
R_{\rm m} = 3.2 \times 10^{8} {M_{1}^{-1/7}} \mu_{30}^{4/7}
\dot{M}_{17}^{-2/7}\; {\rm cm}, 
\label{eq:Rm}
\end{equation}
where $M_{\rm 1}$ is the NS mass in units of 1 M$_{\odot}$,
$\mu_{30}$ is the NS magnetic dipole moment in units of $10^{30}$ G
cm$^{3}$, and $\dot{M}_{17}$ is the mass accretion rate in units of $10^{17}$ g
s$^{-1}$.  

Disk accretion onto a spinning, magnetic NS is expected to be centrifugally
inhibited, once the magnetospheric radius is larger than the
corotation radius, $R_{\rm co}$, and inside the light cylinder radius,
$R_{\rm lc}$ \citep{IS}. The  corotation radius,
$R_{\rm co}$, is the radius at which the Keplerian frequency of the 
orbiting matter is equal to the NS spin frequency: $R_{\rm co} =
(GM/\Omega_{0}^{2})^{1/3} = 1.5 \times 10^{6}{P}_{-3}^{2/3}
(M/{\rm M_\odot})^{1/3}$ cm, where $\Omega_{0}$ is the angular velocity and
$P_{-3}$ is the spin period of the NS in milliseconds.
The light cylinder radius, $R_{\rm lc}$, is the radius at which  
the corotating magnetic field lines reach the speed of light
$c$, $R_{\rm lc}= c/\Omega_{0}$. Outside $R_{\rm lc}$ the magnetic 
field lines cannot corotate any longer and a radiative 
electromagnetic field must be generated. 
If an element of matter $dM_{\rm in}$ is accreted onto a NS from a 
Keplerian disk, part of its energy is stored as rotational
energy of the NS; this can later be used to eject a mass
$dM_{\rm out}$. 
We define the propeller efficiency as the ratio of the ejected
to accreted matter
$\eta_{\rm pro} \equiv dM_{\rm out} / dM_{\rm in}$. 
Moreover we assume that only the rotational energy gained 
by the star during the previous accretion phase can be used 
to eject matter in the propeller phase.

In order to determine the rotational energy gained by the NS 
we first consider the accretion phase. 
During accretion the rate of rotational energy
transferred to the NS can be written in terms 
of the transfer of angular momentum, $L$, at
the inner boundary of the disk:
\begin{equation}
\dot E_{rot} = \Omega_{0}\dot L - \frac{1}{2}\Omega_{0}^{2}\dot I,
\label{torque}
\end{equation}
where $\dot L = I \dot \Omega_{0} + \Omega_{0} \dot I = \dot M l$.
Here $I = \beta^{-1}MR_{\rm NS}^{2}$
is the moment of inertia of the NS, and $l$ 
is the specific angular momentum of the accreted mass.
For NS models with mass above
an initial value of M $\sim$ 1.4 M$_{\odot}$ \citep{Weaver} and 
well below the maximum mass, values of $\beta$ are expected in the 2.3--3.1 
range for a variety of equations of state. 
In our calculations we use $\beta = 2.5$, the value for a uniform 
Newtonian star. 
For a disk truncated at the magnetospheric radius, $R_{\rm m}$, 
we approximate the specific angular momentum transferred 
to the NS with the Keplerian value at $R_{\rm m}$, i.e. 
$l \simeq l_{\rm m} =(GMR_{\rm m})^{1/2}$. According to detailed models of the 
disk-magnetosphere interaction this approximation is accurate in the 
limit in which $R_{\rm m} \ll R_{\rm co}$, whereas it provides an upper limit 
in the case in which $R_{\rm m} \simeq R_{\rm co}$ and non-material (spin-down) 
torques due to the interaction of the NS magnetic field with
disk regions outside $R_{\rm co}$, become important
\citep{GL79,W87,RFS}. 
Since our aim here is to derive an upper limit on the propeller efficiency, we
can safely adopt the  Keplerian value of the specific angular momentum. 
Equation (\ref{torque}) thus writes 
\begin{equation}
dE_{\rm rot} = \Omega_{0} (GMR_{\rm m})^{1/2} dM_{\rm in} -
\frac{1}{2}\beta^{-1}R_{\rm NS}^{2}\Omega_{0}^{2} dM_{\rm in},
\label{key}
\end{equation}
which describes the amount of energy accreted from a
disk ending at the magnetospheric radius. For a given angular
velocity of the NS, the rotational energy, $dE_{\rm rot}$, is highest when the
magnetospheric radius, $R_{\rm m}$, takes the largest value, compatible
with accretion onto the NS surface: that is at $R_{\rm m} = R_{\rm co}$.
Setting $R_{\rm m} = R_{\rm co}$ Eq. (\ref{key}) gives
\begin{equation}
dE_{\rm rot} = \frac{GM}{R_{\rm co}} dM_{\rm in}-\frac{1}{2}\beta^{-1}R_{\rm
  NS}^{2}\Omega_{0}^{2} dM_{\rm in}.
\end{equation}
This is the maximum increase in $E_{\rm rot}$ that an element of
disk-accreting  mass $dM_{\rm in}$ can cause.

\begin{figure}
 \centering
  \includegraphics[width=9.45 cm, angle=-90]{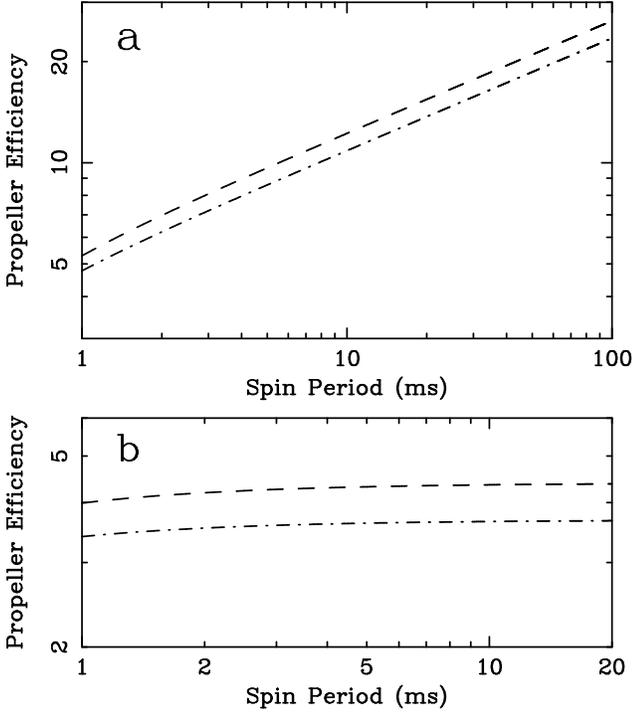}
      \caption{The highest possible propeller efficiency,
      $\eta_{\rm pro}$, the ratio of the ejected mass to the accreted
      mass, versus spin period. The dashed and dot-dashed lines
      correspond to Eq. (\ref{prop1}) (panel a) and Eq. (\ref{prop2}) (panel b), 
      for a NS of 1.4 M$_{\odot}$ and 2.0 M$_{\odot}$, respectively.}.   
         \label{eff}
\end{figure}

We now consider the propeller phase.
According to the virial theorem, the energy required 
to eject a disk mass element, $dM_{\rm out}$, from a radius $R$ to infinity,
is $dE_{\rm eje} = GMdM_{\rm out}/2R$. In order to estimate the maximum
possible propeller efficiency we suppose that ejection 
takes place from the largest possible
radius at which the propeller can work, such that the energy 
required to eject is minimized. That is $R = R_{\rm lc}$.

Setting $dE_{\rm rot} = dE_{\rm eje}$ and assuming that ejection 
takes place at $R_{\rm lc}$, we have for the maximum efficiency of the
propeller: 
\begin{equation}
\eta_{\rm pro} \leq 6.37
\Bigl(\frac{M}{\rm M_{\odot}}\Bigr)^{-1/3} (P_{-3})^{1/3} - 0.57
\Bigl(\frac{M}{\rm M_{\odot}}\Bigr)^{-1} (P_{-3})^{-1}
\label{prop1} 
\end{equation}  
where we used $R_{\rm NS} = 10$ km.

An alternative possibility is that there is no magnetosphere in the
accretion phase, and the disk extends all the way to the NS surface
(such that we set formally $R_{\rm m} \simeq R_{\rm NS}$, where
$R_{\rm NS}$ is the NS
radius). This might be the case in LMXBs which do not display 
coherent pulsations in their persistent X-ray flux.
In this case  
\begin{eqnarray}
\lefteqn{\ \eta_{\rm pro} = \frac{\Omega_{0} (GMR_{\rm
      NS})^{1/2}}{GM/2R_{\rm lc}}-\frac{\beta^{-1}R_{\rm
      NS}^{2}\Omega_0^{2}}{GM/R_{\rm lc}} ={} } \nonumber\\
& & \hspace{0.9cm}{}2\Bigl(\frac{R_{\rm NS}}{R_{\rm
      g}}\Bigr)^{1/2}-\frac{c\beta^{-1}}{GM}R_{\rm NS}^{2}\Omega_0 \ ,   
\end{eqnarray} 
with $R_{\rm g} = GM/c^{2} = 1.48 \times 10^{5} (M/ {\rm M_{\odot}}) $ cm  the
gravitational radius.

\begin{figure}
  \includegraphics[width=6.5 cm, angle=-90]{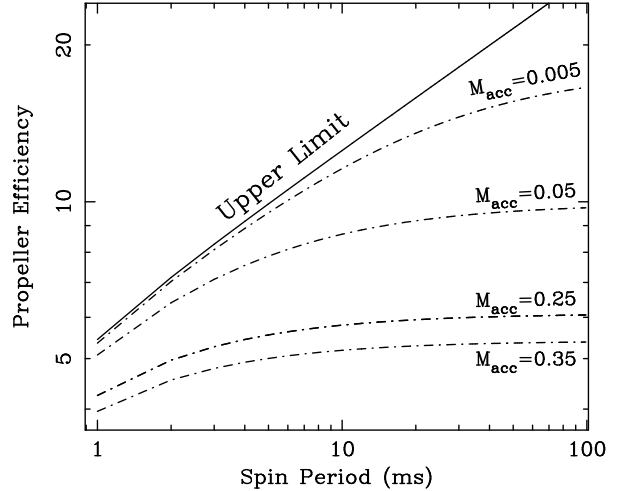}
      \caption{The highest possible propeller efficiency,
      $\eta_{\rm pro}$, versus spin period for different amounts of
      mass accreted onto a 1.3 M$_{\odot}$ NS are shown by dot-dashed lines. 
      The solid line is the propeller efficiency upper limit (Eq. \ref{prop1}).}  
         \label{eff2}
\end{figure} 

For a typical NS, $R_{\rm NS} \simeq 6 R_{\rm g}$, one has
\begin{equation}
\eta_{\rm pro} \leq 5.20\Bigl(\frac{\rm
  M}{M_{\odot}}\Bigr)^{-1/2}-0.57\Bigl(\frac{\rm M}{M_{\odot}}\Bigr)^{-1} (P_{-3})^{-1}
\label{prop2}  
\end{equation}
a result that is similar to that in Eq. (\ref{prop1}).
   
The above derived limits on $\eta_{\rm pro}= dM_{\rm out} / dM_{\rm 
in}$ involve a great deal of fine tuning  
with respect to the ejection of matter from close to the light cylinder. 
Indeed, despite the fact that the NS magnetic field lines approach $c$ at 
$R_{\rm lc}$, it assumed that ejection takes place at much lower speeds, 
precisely the lowest required for matter to reach infinity.   
 
Eqs. (\ref{prop1}) and (\ref{prop2}) are plotted in Fig. \ref{eff} 
as a function of the spin period 
and for different values of the NS mass. 
In the following we discuss the impact of the limits 
in Eqs. (\ref{prop1}) and (\ref{prop2}) on evolutionary models of accreting NS's
binaries and MSP's. 

\section{Discussion}
\label{discussion}

The limits that we have derived on the mass ejection efficiency of the 
propeller mechanism are based on the assumption that, when accretion
takes place, the highest angular momentum is transferred to the NS, whereas 
ejection occurs at the expense of the NS rotational 
energy from the largest radius compatible with the propeller 
mechanism. This minimizes the loss of rotational energy.

Somewhat different limits are obtained depending on whether 
during the accretion phase, the disk ends at the magnetospheric boundary or 
extends all the way to the NS surface. The latter case might be relevant to 
high luminosity NS systems which do not display periodic pulsations. 
The corresponding limit on $\eta_{\rm pro}$ (see Eq. \ref{prop2})
applies only over a limited range of fast NS spin periods. 
Indeed, if the magnetosphere is "squeezed" inside the star
surface during the accretion
phase  (formally $R_m \simeq R_{\rm NS}$), then, as the mass inflow rate
decreases  by up to $\sim$4--5 orders of magnitude \citep[as inferred from
the luminosity NS X-ray transient systems, e.g.,][]{CS00}, the
magnetosphere can expand up to $\sim$20 $R_{\rm  NS}$ at the
maximum. This is due to the stiffness of the NS magnetosphere, reflected in 
the weak dependence of $R_{\rm m}$ on $\dot M$ in Eq. (\ref{eq:Rm}). For the 
NS to be in propeller phase the corotation radius, $R_{\rm co}$, must be 
smaller than the magnetospheric radius $R_{\rm m}$ ($\sim 20 R_{NS}$), 
in turn implying that the NS spin cannot be longer than $\sim30$ ms 
(for NS with spin period longer than this, the propeller efficiency limit in 
Eq. (\ref{prop2}) would thus be inapplicable). In this case, from
Fig. \ref{eff} (panel b), we note that the maximum propeller efficiency 
cannot be larger than $\simeq$ 4 for NS with a spin period of $\simeq1$ ms. 
The limit on $\eta_{pro}$ derived in the case in which there exists a 
magnetosphere also during the accretion phase (see Eq. \ref{prop1})
applies instead over a large range of spin periods and  its value is
$\sim5$, for a spin period of $\simeq$1 ms and increases for longer
periods.

Strictly speaking, our reasoning applies 
only to the accretion/ejection of (infinitesimally) small amounts of 
matter. The limits we derived are thus immediately relevant to NSs
that often cycle between the accretion and ejection regimes.   
These include transient X-ray systems which alternate active periods, lasting 
up to years, to quiescent intervals of up to $\sim100$ yr. During each
outburst cycle the NS paramaters (e.g., mass, spin period, magnetic
dipole moment) change 
only by very small amounts. These systems likely evolve 
through a very long sequence of accretion/ejection cycles, while within 
each cycle the efficiency of the propeller ejection remains limited by  
the values we have derived. It is easy to see that in this case 
the overall efficiency must be limited by the maximum possible value
of $\eta_{\rm pro}$, given by Eq. (\ref{prop1}) independent of whether
accretion  occurs through the magnetosphere or not.

Therefore, flunging away matter with higher efficiency than permitted
by the above limit would require a different ejection mechanism such as the
radio pulsar ejection mechanism \citep{IS}. This mechanism can operate at much
larger radii (comparable to the NS Roche-lobe radius) where the
gravitational potential is much shallower and thus the energy required  
to expel matter to infinity is much smaller \citep[e.g.,][]{b01}.

In the case of accretion and ejection of finite amounts of matter, 
an analitical expression of $\eta_{\rm pro}$ cannot be found. In
fact, in this case, one has to consider the variation of the stellar
parameters during the evolution of the system. The equations relating
the rotational energy acquired during accretion to the ejection
energy lost during the propeller phase depend on the instantaneous
value of the stellar mass and angular velocity, and their evolution
can be followed numerically. To obtain an estimate of the maximum
propeller efficiency we again assumed that the accretion phase takes
place from the corotation radius, the ejection phase from $R_{\rm lc}$
and that all the rotational energy acquired during accretion is used to 
eject matter from the system. 
In our numerical integrations we neglected the decrease in the 
NS radius which results from increasing its mass through accretion.
In any case, using the simple relation $R_{NS}\propto M_{NS}^{-1/3}$ 
we checked that this is a small effect and the induced variations on $\eta_{\rm
  pro}$ are less that $\sim2\%$. 
The parameters that we used in our calculations are those in the
evolutionary models of \citet{ES} and \citet{TS}. In the first paper the
authors assume that propeller ejection
takes place from an increasing radius, as the mass inflow rate toward
the NS decreases.  
Using their parameters \citep{ES}, i.e. NS mass of
$\sim1.4$ M$_{\odot}$, initial spin period $\sim0.03$ s, 
$\sim0.25$ M$_{\odot}$ accreted onto the NS, and a final spin period of
$\sim 3.6$ ms, our calculations show that the
efficiency is bound by the maximum value of $\lesssim5$.
This is a resonable result since accretion causes $\Omega_{0}$ to increase 
and consequently, ejection
takes place (mainly) at smaller values of  $R_{\rm lc}$. In this case the
energy required to eject matter is higher and the  propeller less
efficient. An additional effect comes from the fact that the
corotation radius slightly increases during accretion due to the building up of the NS mass.
However, our calculations show that this effect is
small (due to the limited range in NS mass). Infact in all cases we obtained 
a lower efficiency than the maximum value of $\eta_{\rm
  pro}$ of Eq. (\ref{prop1}), which holds for short accretion/ejection cycles.
Fig. \ref{eff2} shows that the upper limit of Eq. (\ref{prop1})
is approached only when the accreted mass on the NS is very small.  
We note that the propeller efficiency required in the work of
\citet{ES} is $\sim2.6$, consistent with the limit we have
derived. 
It is more difficult to carry out a comparison with the work of
\citet{TS}. We refer to their models for relatively close
binary systems with light donors ($P_{\rm orb}< 10$ d, $M_{\rm donor}
< 1.3$ M$_{\odot}$). This is because, in those cases, the
mass-transfer rate remains sub-Eddington during the entire binary
system evolution, preventing the formation of super-Eddington strong
outflows and causing the binary system to eject matter mainly via the
propeller effect. We use a NS mass of $\sim$1.3
M$_{\odot}$, $\sim$0.1--0.4 M$_{\odot}$ accreted onto the NS and an
initial NS spin period in the range $\sim0.001$--$0.006$ s. Again we
find the required propeller efficiency in Tauris
\& Savonije (in some cases $\eta_{\rm pro}\lesssim1$) 
is smaller than our numerically estimated upper limit ($\eta_{\rm pro}\lesssim7$).

A refined treatment of the propeller effect for a magnetic accreting ms pulsar
was carried out by \citet{ug}. Using a magnetohydrodynamic approach, and 
considering a detailed model for the interaction between the NS
magnetosphere and the accretion disk, they showed that propeller outflows can 
be expected from a fast rotating ($P_{\rm spin}\sim 1.2$--$1.6$~ms) weakly magnetic NS 
($B\sim5\times10^{8}$--$2\times10^{9}$~G).
A comparison between 
their results and ours $\eta_{\rm pro}$ is straightforward
if one consider that all the NS rotational energy used in the propeller phase
derives from accretion of matter 
at the corotation radius.
Dividing their total ejected matter ($\dot{M}_{j}+\dot{M}_{w}$) by
the above estimated accreted matter, we obtain $\eta_{\rm pro}\sim0.4$, 
for a $1.26$ ms pulsar with $B\sim10^9$ G.

In summary the evolutionary calculation of \citet{ES} and \citet{TS} 
correspond to a propeller efficiency of $\sim$1--3, which 
as such would be adequate to explain how the 
ejection of a large fraction of the mass transferred 
by the companion can take place in these systems. 
The upper limit that we have derived in this paper for the 
propeller efficiency is only a factor of $\sim$2 higher. 
However achieving such a limit would require a high degree 
of fine tuning. This indicates that a propeller efficiency 
of order $\sim$2--3 might be difficult to achieve in real systems. 

As suggested by the deteiled propeller simulations by \citet{ug} 
a propeller efficiency $<1$ might be more realistic. 
If this was the case, then a different mechanism to 
eject from the binary most of the mass transferred 
by the companion is required. The onset of the radio pulsar 
ejection mechanism is thus probably to be favored.

\begin{acknowledgements} 
MF acknowledges the French Space Agency  (CNES).  
EB acknowledges support from ASI (I/023/05/0) and from the 
Italian Minister of University and Technological Research 
through grant PRIN2004023189. EB thanks G. Lavagetto 
for usefull discussions and CEA Saclay, DSM/DAPNIA/SAp for 
hospitality during part of this work.
\end{acknowledgements}

\end{document}